\documentclass[aps,twocolumn, showpacs]{revtex4}
\usepackage{float}
\usepackage{amsmath}
\usepackage{graphicx}
\usepackage{hyperref}
\usepackage{color}
\usepackage{amssymb}
\usepackage{amsfonts}
\usepackage{lipsum,appendix}
\usepackage[makeroom]{cancel}
\newcommand{\be}{\begin{equation}} 
\newcommand{\ee}{\end{equation}} 
\newcommand{\bea}{\begin{eqnarray}} 
\newcommand{\eea}{\end{eqnarray}} 
\newcommand{\bqa}{\begin{eqnarray}}
\newcommand{\eqa}{\end{eqnarray}}
\newcommand{\bwt}{\begin{widetext}}
\newcommand{\ewt}{\end{widetext}}
\newcommand{\mb}{\mathbf}
\newcommand{\mc}{\mathcal}
\newcommand{\nn}{\nonumber \\}
\newcommand{\sig}{\sigma}

\newcommand{\e}{\epsilon}

\newcommand{\w}{\omega}

\newcommand{\figImMT}
{\begin{figure}[htbp]
        \centering
        \includegraphics[angle=270,width=7cm]{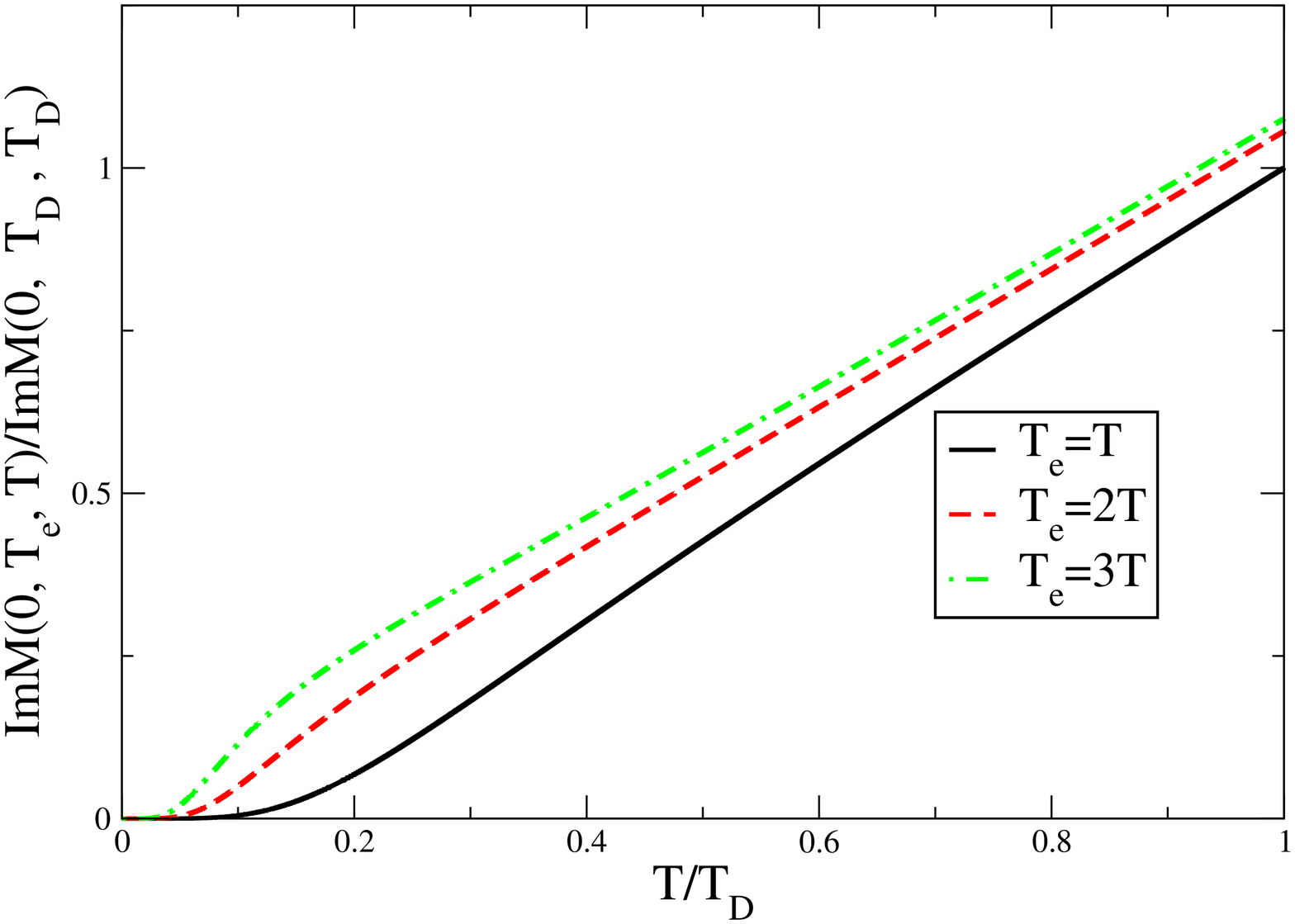}
           \caption{Variation of the imaginary part of the memory function with temperature at zero frequency at different electron temperatures. Here $M''(0,T_e, T)$ is scaled with $M''(0,T_e=T_D, T=T_D)$ and the temperature is scaled with the Debye Temperature.}
           \label{fig:figimmt}
	\end{figure}
} 
\newcommand{\figImM}
{\begin{figure}[htbp]
        \centering
        \includegraphics[angle=270,width=7cm]{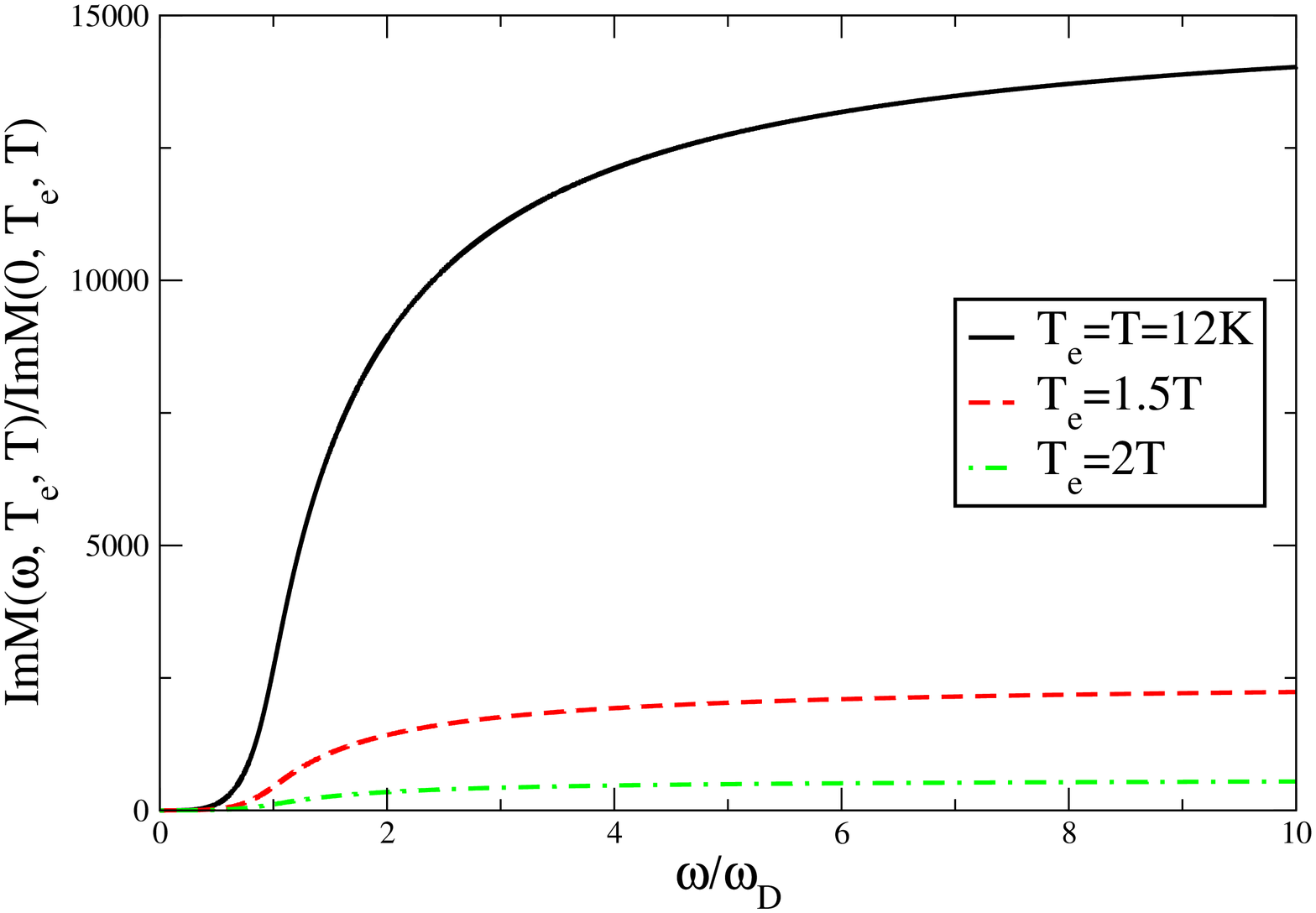}
           \caption{Variation of the imaginary part of the memory function with frequency at two phonon temperatures and various different electron temperatures. Here $M''(\w,T_e, T)$ is scaled with $M''(0,T_e, T)$ and the frequency is scaled with the Debye frequency.}
           \label{fig:figimm}
	\end{figure}
} 
\newcommand{\figImMmT}
{\begin{figure}[htbp]
        \centering
         \includegraphics[angle=270,width=7cm]{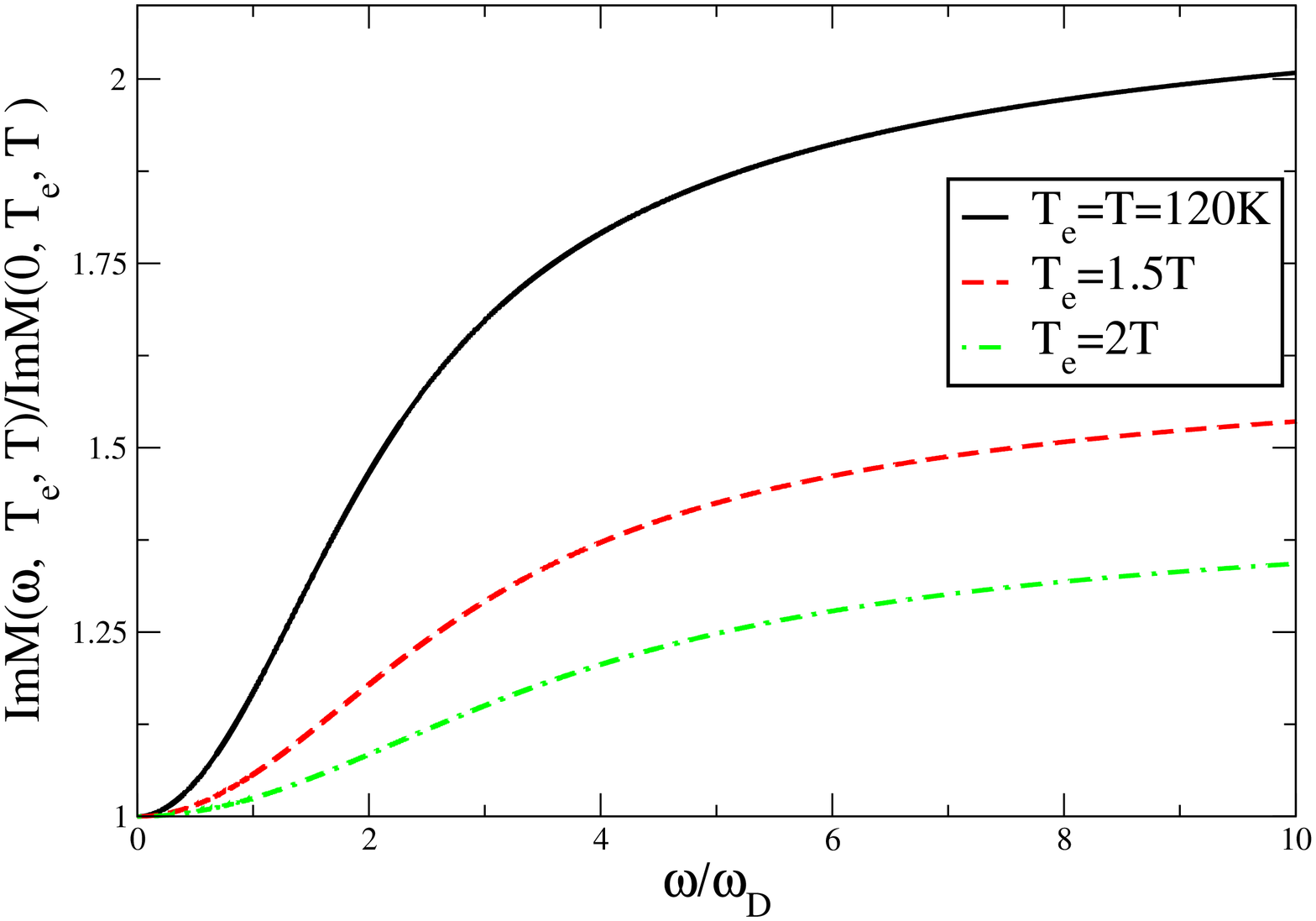}
           \caption{Variation of the imaginary part of the memory function with frequency at an intermediate phonon temperature, $120K$ and various different electron temperatures. Here $M''(\w,T_e, T)$ is scaled with $M''(0,T_e, T)$ and the frequency is scaled with the Debye frequency.}
           \label{fig:figimm-mT}
	\end{figure}
} 
\newcommand{\figImMHT}
{\begin{figure}[htbp]
        \centering
         \includegraphics[angle=270,width=7cm]{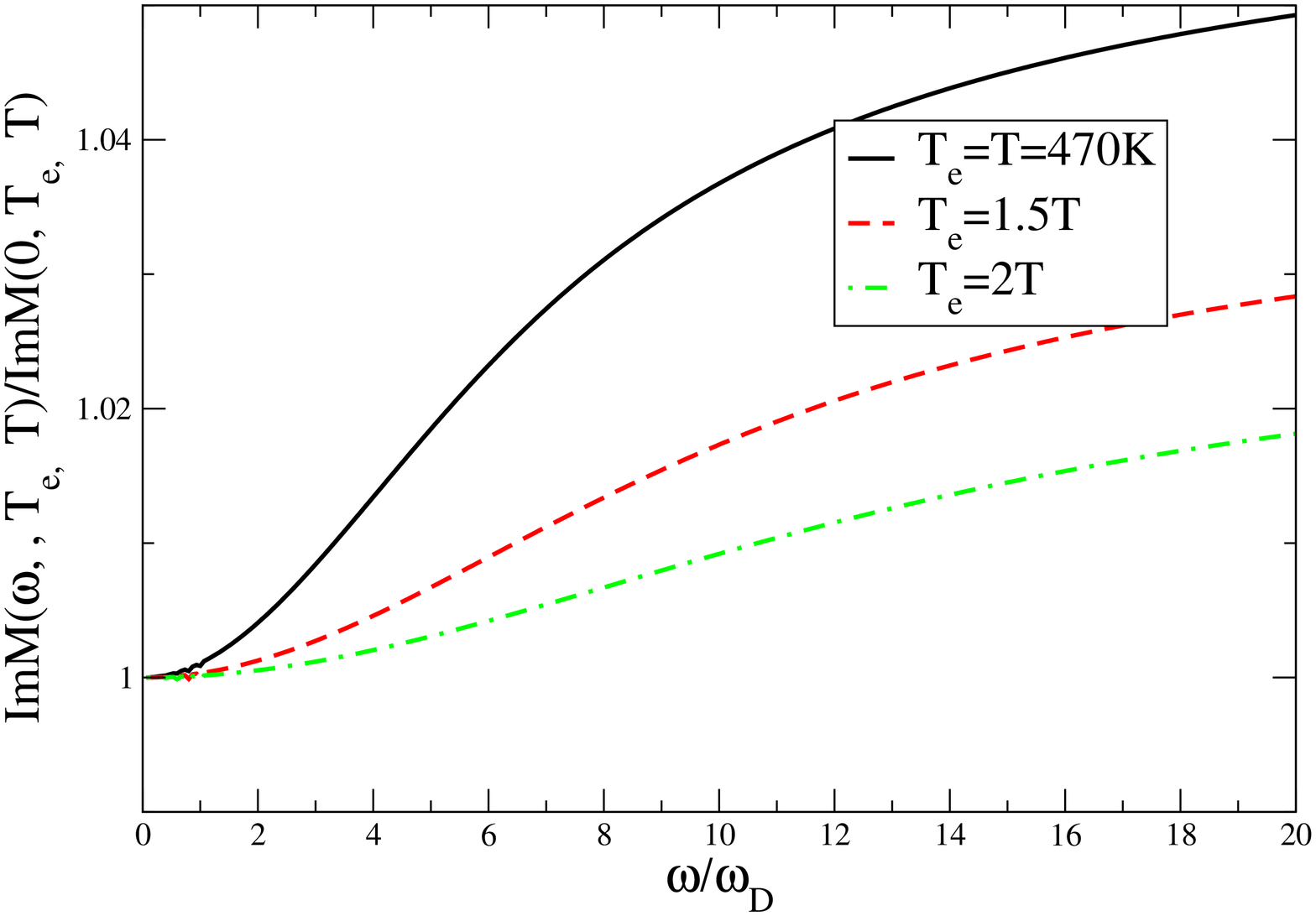}
           \caption{Variation of the imaginary part of the memory function with frequency. The plot is shown at a phonon temperature $= 470$K higher than the Debye temperature $\sim 350$K and various different electron temperatures. Here $M''(\w,T_e, T)$ is scaled with $M''(0,T_e, T)$ and the frequency is scaled with the Debye frequency.}
           \label{fig:figimm-HT}
	\end{figure}
} 

\begin{document} 
\baselineskip 12pt 
\title {Hot-electron relaxation in metals within  the G\"otze-W\"olfle memory function formalism }
\author{Nabyendu Das and Navinder Singh }
\affiliation{Theoretical Physics Division, Physical Research Laboratory, Ahmedabad-380009, India} 
\date{\today}
\begin{abstract}
We consider non-equilibrium relaxation of electrons due to their coupling with phonons in a simple metal. In our model electrons are living at a higher  temperature than that of the phonon bath, mimicking a non-equilibrium steady state situation. We study the relaxation of such {\it hot electrons} proposing a suitable generalization of the memory function formalism formulated by G\"otze and W\"olfle\cite{GW}. We derive analytical expressions for both dc and optical scattering rates in various temperature and frequency regimes. Limiting cases are in accord with the previous studies. An interesting feature, that the dc scattering rate at high temperatures and optical scattering rate at high frequencies, are independent of the temperature difference between the electrons and the phonons is found in this study. The present formalism forms a basis which can also be extended to study hot-electron relaxation  in more complex situations. 
\end{abstract}
\maketitle 
\section{Introduction}
 In electronic systems conductivity is one of the most important quantity to study. It is related to the scattering rate of the charge carriers due to various interactions. Theoretical studies of electron transport in a solid suffer mainly from two difficulties. One is related to the proper choice of a scattering rate or the inverse quasi-particle lifetime and the subtle difference in scattering rates found from a transport measurement, e.g. resistivity measurement to the one found in a spectroscopic measurement such as ARPES\cite{Mahan}. Other is to incorporate the  non-equilibrium nature of the charge carriers in certain transport measurements. In an equilibrium transport calculation, Drude approximation is the first starting point which treats electrons as non-interacting classical particles and an average life time or collision time is assigned for all. It relates the frequency dependent conductivity $\sigma(\w)=\frac{\sigma_{DC}}{1-i\w\tau},\,\,\sigma_{DC}=\frac{ne^2\tau}{m}$.   Here $n$, $e$, $m$ and $\tau$ are the density, the charge, the mass or more precisely the effective mass and the lifetime of the electrons respectively. From the expression of Drude conductivity we see that the electronic lifetime or the inverse scattering rate and the effective mass are the most important parameters to determine the transport properties of an electronic system. Among them, change in the $\tau$ becomes more prominent in presence of various interactions and the studies of the transport properties becomes synonymous to the studies of the scattering rates.  In the Drude theory, the collision time $(\tau)$ is estimated as the ratio of average inter-particle separation$(l)$ and the average velocity$(v_{avg})$. Following equipartition theorem, it is assumed $v^2_{avg}\sim k_BT$, the thermal energy. It undermines electron velocity and hence the scattering rates in a metal. Later the Fermionic nature of the electrons was considered and $v_{avg}$ was replaced by the $v_{F}$, the Fermi velocity by Sommerfield\cite{Dressel}. Next comes Boltzmann equation approach. In this approach, instead of looking at individual particle dynamics one considers the time evolution of the distribution function for the collection of particles under external perturbations\cite{Ashcroft}. It is certainly a major improvement over the Drude-Sommerfield approach.  However this approach is mostly limited to semi-classical approximations. Also the related equations governing the time evolution of the distribution function is  mostly solved within relaxation time approximation. The later is confined to near equilibrium situations and often uses frequency dependent quasi-particle lifetime for better results. All these traditional approaches are restricted in many aspects and thus fail to explain transport behavior in many physical systems. The time scale enters into the transport calculations should be the transport life time, which is often misunderstood as the scattering life time. They are certainly different. The transport properties are defined and hence measured in the presence of an external field put in a preferred direction. Thus unlike the case of equilibrium spectroscopic properties, scattering in all the directions can not have equal contributions in determining transport coefficients. For example in a resistivity measurement in certain direction, scattering of charged particle in the opposite direction or the back scattering should have the dominating contribution.  This difficulty can be overcome by incorporating vertex correction in quantum many body theory which results in the famous $(1-\cos\theta_{\mb{k}\mb{k'}})$ contribution to the resistivity expression in an electron-phonon system\cite{Mahan}. Here $\theta_{\mb{k}\mb{k'}}$ is the angle between the incident and the scattered direction of an electron. However since transport properties like conductivity, are related to the two particle correlations, it is never fully justified to express them in terms of the single particle self-energy and vertex corrections.  The transport calculations becomes more problematic, in case of strong electron-electron interactions and or in the non-equilibrium situations.\\ \indent
 Here comes the importance of the memory function approach to this problem. This approach was originally introduced by Mori\cite{Mori} and Zwanzig \cite{Zwanzig} separately in studying non-equilibrium problems. A detailed review of the original work can be found in Ref. \cite{Forster, Fulde, Vladimirov}. Later it was successfully introduced in calculating electrical conductivity by  G\"otze and  W\"olfle\cite{GW}. Similar approaches are also used recently to understand the transport properties of various metalic systems \cite{Lucas, Sachdev, Patel}. They express memory function for electronic transport in terms of the two particle correlations and calculate it at the lowest order in electron phonon coupling. In their approach corrections beyond the single particle self energy calculations and the vertex correction, at least in the lowest order are incorporated naturally.\\ \indent
 Now let us focus on some situations where adopting an equilibrium description of electronic conduction where both the electron and phonon temperatures are same is not justified. In a situation where large current flows through a sample in presence of a large external electric field or strong pulse\cite{Eesley, Hirori, Rogier} the current-voltage relation becomes non-ohomic or non linear. In such situations, within an experimentally accessible time scale the system remains in a non-equilibrium state. This occurs due to the following reason. A flow of large current heats the system. Due to the lower specific heat of electrons than that of the phonons, the electron temperature becomes higher.  The electrons owing to their shorter relaxation time scale (due to the electron-electron interactions) equilibrate much faster than the phonon subsystem. Thus they end up at a quasi-equilibrium state where their effective temperature is higher than the phonons. They are termed as ``hot-electrons'' here. They remain hot until they relax due to their coupling with phonons. We can also form such a quasi-equilibrium steady state using continuous wave laser excitations.  In this situation electron-phonon scattering rate can not be calculated within an equilibrium description.  One needs to consider ``hotness'' of electrons in a suitable way. This motivates Kaganov et. al.\cite{KLT} to propose the so called Two Temperature Model(TTM) which was later described in a modern language by Allen\cite{Allen}. Detailed discussions on  this model can be found in a recent review by one of the authors\cite{Navinder}. The relaxation of hot-electrons
 due to coupling with phonons is also discussed in many recent literature in a different contexts within different approaches\cite{Ma, Kristen, Sds, Vidmer,Fatti, Chaturangi, Huang}. Adaptation of such non-equilibrium picture got more impetus with the advent of pump-probe spectroscopy.  Later is recently  being used to study important quantities like electron-phonon coupling in cuprates and other correlated systems\cite{Zhang}.

The combined effects of these two aspects, even in a simple metal are not well understood. Thus we consider a simple metal in a non-equilibrium or quasi-equilibrium situation where the electron subsystem is excited to a higher temperature than that of  the phonon bath 
and attempt a memory function generalization of the TTM formalism to calculate the effects of the electron-phonon coupling on the electrical conductivity. This is a new formalism which includes transport aspects in a non-equilibrium situation. We derive analytical expressions for both dc and optical scattering rates in various temperature and frequency regimes. Limiting cases are in accord with Bloch-Boltzmann formula as found in the previous studies\cite{GW}. An interesting feature, that the dc scattering rate at high temperatures and optical scattering rate at high frequencies, are independent of the temperature difference between the electrons and the phonons is also a major outcome of our approach. We also discuss the possibilities of extending this approach to many other directions.

This paper is organized as follows.  In section \ref{sec:memory} first we review the memory function Formalism used by G\"otze and W\"olfle(GW) to calculated electrical
conductivity in the subsection \ref{subsec:GW}. Then in the subsection \ref{subsec:NGW} we propose a new generalization of it in case of Hot-electron systems and derive an expression for scattering rate due to electron-phonon interactions. In the next section \ref{sec:result} we show our results. Here we analytically calculate the electron-phonon scattering rate in various limiting cases also show it's full behavior using numerical calculations. Finally in section \ref{sec:discussion} we summarize our results and present our conclusions.
\section{Memory function Formalism}
\label{sec:memory}
Our aim is to study the electron relaxation in a non-equilibrium steady state situation and to calculate the scattering rate using the Memory function formalism. First we review the later in case of electron-phonon interaction in metals in equilibrium. Then in the next subsection we develop a generalization of the formalism in case of a non-equilibrium steady state situation. 
\subsection{G\"otze and W\"olfle(GW) formalism\cite{GW} for electron-phonon interactions} 
\label{subsec:GW}
We consider a degenerate gas of electrons having isotropic energy dispersion $\epsilon_k=\frac{k^2}{2m}$, where $k$ is the momentum and $m$ is the electronic mass. Also in the whole discussion we follow the system of units where $\hbar=1$ and $k_B=1$. We start with the following Hamiltonian describing electron-phonon interactions,
\bea 
H&=& H_0+H_{ph}+H'\nn
H&=&\sum_{\mb{k}} \epsilon_k c^\dagger_\mb{k}c_\mb{k}+\sum_\mb{q} \w_q\left(b^\dagger_\mb{q} b_\mb{q} +\frac{1}{2}\right) \nn
&& +\sum_{\mb{k},\, \mb{k'}}\left[ D(\mb{k}-\mb{k'})c^\dagger_{\mb{k}\sigma}c_{\mb{k'}\sigma}b_{\mb{k}-\mb{k'}} +H.\, c.\right].
\label{eq:H}
\eea 
 The first and the second parts of the Hamiltonian represent free electrons and free phonons respectively. The electron-phonon interaction is depicted by the last term, where the electron-phonon matrix elements is given as,
\bea 
D(\mb{q})=(1/\sqrt{2m_iN\w_q}) q C_q.
\label{eq:e-ph-mat}
\eea 
The phonons are considered as acoustic having a dispersion of the form $\w_q=cq$. Here $c$ is the sound velocity and $q$ is the phonon momentum. The constant $C_q=1/\rho_F$ for metals. This completes the Hamiltonian description of  system considered here. Before using the memory function appraoch to a two temperature scenario, let us make a summary of the previous work by  G\"otze and  W\"olfle\cite{GW}.
In a theory of transport properties, we look for the linear response of a physical quantity represented by an operator $A$ due to an external perturbation coupled to another operator $B$. The response can be quantified and can be expressed in terms of the correlation function of these two operators as,
 \bea 
 \chi_{AB}(z)=\langle\langle A;B\rangle\rangle=-i\int_0^\infty e^{izt}\langle [A(t), B(0)]\rangle dt.
 \eea 
Here $<......>$ describes expectation value of the inside operator and $<<.......>>$ describes Laplace transform of the expectation value of the same.  Being a correlation function, $\chi_{AB}(z)$ must maintain causality and must be compatible with the equation of motions  obeyed by the operators and thus it has  the following  properties,
 \bea
 \chi_{AB}(z)=\langle[A,B]\rangle /z \,\,{\rm for}\,\, |z|\rightarrow \infty.
 \eea 
  In terms of the Laplace transform of derivative, equation of motion is given as,
 \bea 
z\langle \langle A;B\rangle \rangle &=& \langle[A,B]\rangle +\langle\langle[A,H];B\rangle \rangle\nn
&=& \langle[A,B]\rangle - \langle\langle A;[B,H]\rangle \rangle.
\label{eq:eqm}
 \eea 
 Since $\chi_{AB}(z)$ is analytic, we can write it as a spectral decomposition as follows,
 \bea
 \chi_{AB}(z)=\frac{1}{\pi}\int d\w \chi_{AB}^{\prime\prime}(\w)/(\w-z).
 \eea 
  Our focus is on frequency dependent conductivity. It is related to the current-current correlation function within Kubo formalism\cite{Kubo} as,
 \bea 
 \sigma(z)=-i\frac{e^2}{z}\chi(z)+i\frac{\w_p^2}{4\pi z}.
 \label{eq:sig0}
 \eea 
 In this case correlation is considered between the components of currents in the different directions. In an isotropic case, the related correlation is defined as,
 \bea
   \langle\langle j_i;j_j\rangle \rangle=-\delta_{ij}\chi(z), \,\,\, j_i=\sum v_i(\mb{k}) c^\dagger_{\mb{k}\sigma}c_{\mb{k}\sigma}.
 \eea
  Now we attempt to express the frequency dependent conductivity in terms of a {\it Memory function} $M(z)$. For this problem the later is defined by the relation 
  \be 
  \sigma(z)=\frac{i}{4\pi}\w_p^2/[z+M(z)],
  \ee
 where $\w^2_p=\frac{ne^2}{m}$ is the square of the plasma frequency. Using Eqn.\ref{eq:sig0}, $M(z)$ can be written as, 
 \bea 
 M(z)=\frac{z\chi(z)}{\chi_0-\chi(z)}. 
 \label{eq:mz} 
 \eea 
 Since the memory function is related to the correlation function, it has the following properties, 
 \bea 
 M(z\rightarrow \infty)&\rightarrow& 1/z, \,\, 
 M^*(z)=M(z^*),\nn
 M^\prime(\w)&=&-M^\prime(-\w), \,\,\, M^{\prime\prime}(\w)=M^{\prime\prime}(-\w), 
 \eea 
 and has a spectral decomposed form,
 \bea
 M(z)=\frac{1}{\pi}\int d\w M^{\prime\prime}(\w)/(\w-z).
 \eea
For vanishing impurity concentrations and electron-phonon coupling, the current $j$ is nearly conserved and $\chi, M\rightarrow 0$ in this limit. Thus neglecting the terms $\mc{O}(\chi M)$ in Eqn.\ref{eq:mz}, we get,
 \be
 z\chi(z) =\chi_0 M(z) .
 \ee
 Substituting $A=B=j_1$ in equation of motion (Eqn.\ref{eq:eqm}) we get,
 \be
 z\langle\langle j_1;j_1\rangle\rangle=\langle\langle [j_1,H^\prime];j_1\rangle\rangle.
 \ee
  Now considering $A=[j_1,H^\prime]$, and $B=j_1$ we get the following relations,
 \be 
 z\langle\langle A;j_1\rangle\rangle=\langle [A,j_1]\rangle - \langle\langle A;A\rangle\rangle.
 \ee 
 Since the first term in the right hand side is frequency independent, we get in the $z=0$ case,
 \be
\langle [A,j_1]\rangle=\langle\langle A;A\rangle\rangle_{z=0}.
 \ee 
 Using the above relation, we can rewrite the memory function as,
 \bea 
 \chi_0 M(z)=\left(\langle\langle A;A\rangle\rangle_{z}-\langle\langle A;A\rangle\rangle_{z=0}\right)/z.
 \eea 
 Defining $\phi(z)=\langle\langle A;A\rangle\rangle_{z}$, where $ A=[j_1, H^\prime]$,  the memory function can be approximated in the lowest order in electron-phonon coupling as,
 \bea 
 M(z)\approx \left[\phi(z)-\phi(0)\right]/(z\chi_0).
 \label{eq:Mphi}
 \eea  
 The above expression forms the basis for calculating the memory function for different interactions in the GW original work. However our focus is on electron-phonon interaction and in this case $M(z)$ can be calculated as follows.
 
 We use the expression for electrical current as $\mb{j}=\sum_{\mb{k}\sig}\mb{v(k)}c^\dagger_{\mb{k}\sigma} c_{\mb{k'}\sigma}$, where $\mb{v(k)}$ is the velocity of the carriers. Now from the electron-phonon interaction part of the Hamiltonian (Eqn.\ref{eq:H})we get, 
 \be
 A= \sum[v(\mb{k})-v(\mb{k'})][D(\mb{k}-\mb{k'})c^\dagger_{\mb{k}\sigma} c_{\mb{k'}\sigma}b_{\mb{k}-\mb{k'}}-H.c.].
 \ee
 \bwt
Using the above expression, the two particle correlator $\phi(z)$ can be written in the form,
 \bea
 \phi(z)&=&\sum_{kk'\sig}\sum_{pp'\tau}[v(\mb{k})-v(\mb{k'})][v(\mb{p})-v(\mb{p'})]\nn
 &&\times\left[[D(\mb{k}-\mb{k'})][D^*(\mb{p}-\mb{p'})] \langle\langle c^\dagger_{\mb{k}\sigma} c_{\mb{k'}\sigma}b_{\mb{k}-\mb{k'}}; c^\dagger_{\mb{p}\tau}c_{\mb{p'}\tau}b^\dagger_{\mb{p}-\mb{p'}}\rangle\rangle+ c.c.\right]\nn
 &=&-\frac{2}{3}(m)^{-2}\sum_{kk'}\left|D(\mb{k}-\mb{k'})\right|^2(\mb{k}-\mb{k'})^2[f(1-f')(1+n)-(1-f)f'n]\nn
 &&\times\left[\frac{1}{\epsilon_k-\epsilon_{k'}-\omega_{\mb{k}-\mb{k'}}+z}+\frac{1}{\epsilon_k-\epsilon_{k'}-\omega_{\mb{k}-\mb{k'}}-z}\right].
 \label{eq:phi}
 \eea 
Here $f=f(\epsilon_k, T)$, $f'=f(\epsilon_{k'}, T)$ are the Fermi-Dirac distribution functions at different energies and $n=n(\w_q, T)$ is the Bose-Einstein distribution function at  temperature $T$. Thus we reach at the G\"otze and W\"olfle expression for the correlator(Eqn. (53a), Ref.\cite{GW}),
\bea 
 \phi(z)&=&\frac{2}{3}(m)^{-2}\sum_{kk'}\left|D(\mb{k}-\mb{k'})\right|^2(\mb{k}-\mb{k'})^2[f(1-f')(1+n)-(1-f)f'n]\nn
 &&\times\left[\frac{1}{\epsilon_k-\epsilon_{k'}-\omega_{\mb{k}-\mb{k'}}+z}+\frac{1}{\epsilon_k-\epsilon_{k'}-\omega_{\mb{k}-\mb{k'}}-z}\right].
\eea 
Using the above expression in Eqn.\ref{eq:Mphi}, followed by analytic continuation and after some algebra, the imaginary part of the memory function turns out to be,
\bea
M''(\w)&=& \frac{2}{3}\frac{\pi}{m^2\chi_0}\sum_{kk'} \left|D(\mb{k}-\mb{k'})\right|^2(\mb{k}-\mb{k'})^2[f(1-f')(1+n)-(1-f)f'n]\nn
&&\times\frac{1}{\w}\left[\delta(\epsilon_k-\epsilon_{k'}-\omega_{\mb{k}-\mb{k'}}+\w)-\delta({\epsilon_k-\epsilon_{k'}-\omega_{\mb{k}-\mb{k'}}-\w})\right].
\label{eq:ImM}
\eea
These summarizes the memory function formalism to the case of electron-phonon coupling and we will extend the  above expression in a two temperature scenario in the next subsection.
 
\subsection{Extension of the GW formalism to a non-equilibrium steady state situation}
\label{subsec:NGW}
Now we make a generalization of the above expression to the case of TTM in a phenomenological manner. Here we consider the case when the temperature of the electron subsystem and the phonon subsystem are different and are given by $T_e=1/\beta_e$ and $T=1/\beta$ respectively. Thus the thermal factor before the first delta function in Eqn.\ref{eq:ImM} becomes,
\bea 
\frac{1}{\w}\frac{1}{e^{\beta_e\epsilon_k}+1}\frac{e^{\beta_e\epsilon_{k'}}}{e^{\beta_e\epsilon_{k'}}+1}\frac{e^{\beta(\epsilon_k-\epsilon_{k'}+\w) }}{e^{\beta(\epsilon_k-\epsilon_{k'}+\w)}-1}
-\frac{1}{\w}\frac{e^{\beta_e\epsilon_{k}}}{e^{\beta_e\epsilon_k}+1}\frac{1}{e^{\beta_e\epsilon_{k'}}+1}\frac{1}{e^{\beta(\epsilon_k-\epsilon_{k'}+\w)}-1}.
\eea 
The thermal factor before the another delta function has the same structure but with $\w\rightarrow -\w$. In the expression for the imaginary part of the memory function, the combination of the thermal factor and the delta comes as,
\bea 
f_T=&&\frac{1}{\w}e^{\beta_e\epsilon_{k'}}\delta(\epsilon_k-\epsilon_{k'}-\omega_{\mb{k}-\mb{k'}}+\w)\frac{1}{e^{\beta_e\epsilon_k}+1}\frac{1}{e^{\beta_e\epsilon_{k'}}+1}\frac{1}{e^{\beta(\epsilon_k-\epsilon_{k'}+\w)}-1}\left(e^{\beta(\epsilon_k-\epsilon_{k'}+\w)}-e^{\beta_e(\epsilon_{k}-\epsilon_{k'})}\right)\nn
&& + {\rm terms\,\, with\,\,}\w\rightarrow -\w.
\eea 
To simplify the Eqn.\ref{eq:ImM} further, we insert an integrated delta function ($1=\int dq\delta(q-\left|\mb{k}-\mb{k'}\right|)$) and the resulting expression becomes,
\bea
M''(\w)&=&\frac{2}{3}\frac{\pi N^2}{m^2\chi_0}\int4\pi\frac{k^2dk}{8\pi^3}\int2\pi\frac{k'^2dk'}{8\pi^3}\int_0^\pi d\theta\sin\theta\frac{1}{2m_{i}N\omega_{\mb{k}-\mb{k'}}}\left|\mb{k}-\mb{k'}\right|^4C^2_{\mb{k}-\mb{k'}}\times\int dq\delta(q-\left|\mb{k}-\mb{k'}\right|) \times f_T.
\eea
In case of metals, three of the integrals can be preformed exactly and the expression \ref{eq:ImM} becomes,
\bea
M''(\w)
 &=& M_0\int_0^{q_D} dq q^4\left\{ \frac{1}{\w}\left(\left(n(\beta,\w_q)-n(\beta_e, \w_q-\w)\right)  ( \w-\w_q)\right)
 + \w\rightarrow -\w\right \}.
 \label{eq:ImMf}
\eea
 The details of the calculations are given in Appendix \ref{sec:caln}. Here $n(\beta,\e)$ is the Bose-Einstein distribution obeyed by the phonons living at a temperature $1/\beta$ and having energy $\e$.
Above relation is the general result for the imaginary part of the memory function in a coupled electron-phonon system when electron variables are integrated out. It describes the scattering rate of the electrons coupled to the phonon bath when they are at two different temperatures.  We will discuss the above expression in both the dc and frequency dependent cases in the next section. We need to perform the integral numerically to get the full behavior in arbitrary temperature and frequency regime. However in various limiting cases, analytical results can be obtained and are discussed as follows.
\section{Results}
\label{sec:result}
 Case-I: First we consider the low frequency limit. In this case Eqn.\ref{eq:ImMf} can be simplified as follows.
 \bea
 M''(\w<<\w_D)
 &=&\lim_{\w\rightarrow 0}M_0\int_0^{q_D} dq q^4\left\{ \frac{1}{\w}\left(\left(n(\beta,\w_q)-n(\beta_e,\w_q-\w)\right)  ( \w-\w_q)\right)
 + \w\rightarrow -\w\right \}\nn
 &=& M_0\int_0^{q_D} dq q^4\left\{ \left(\left(n(\beta,\w_q)-n(\beta_e,\w_q)+\w n'(\beta_e,\w_q)-\w^2 n''(\beta_e,\w_q)+\w^3 n'''(\beta_e,\w_q)+\mc{O}(\w^4)\right)  \left( 1-\frac{\w_q}{\w}\right)\right) \right.\nn
 &&\left. + \w\rightarrow-\w\right \}\nn
 &=& 2M_0\int_0^{q_D} dq q^4 \left(n(\beta,\w_q)-n(\beta_e,\w_q)-\w_q n'(\beta_e,\w_q)-\w^2 n''(\beta_e,\w_q)-\w^2\w_q n'''(\beta_e,\w_q) \right) .
 \label{eq:case1}
 \eea 
The frequency derivative of the Bose-Einstein distribution, $n'(\w_q-\w)=\frac{\partial n(\w_q-\w)}{\partial (\w_q-\w)}$ and its higher derivatives appearing here can be written in terms of the Bose function itself and is given in the Appendix \ref{sec:dn}. 
 In the above calculations, integrals can be written in terms of the dimensionless variables. It is done by scaling the momentum $q$  by the temperature scales defined by $T_e \,{\rm or}\, T$ as suitable. When both the electron and phonon temperatures are lower than Debye temperature, the upper limit in the integral for the dimensionless variable becomes infinite. The resulting integrals $(D_n(x))$ are known as Debye integrals, as they appeared first in the specific heat calculations within the Debye model. They are related to the Gamma function and the Riemann Zeta function and are shown in the Appendix \ref{sec:d-func}. 
 Now we discuss the above equation in different sub cases.\\
 Subcase (a): We first consider $\w=0$ and $T, T_e<<\w_D$, i.e when both the electron temperature and the phonon temperature are lower than the Debye frequency. In this case,
 \bea
 M''(\w= 0)&=&  2M_0\left( \frac{1}{(c\beta)^5} \Gamma(5)\zeta(5) -\frac{1}{(c\beta_e)^5} \Gamma(5)\zeta(5)+\frac{1}{(c\beta_e)^5} \Gamma(6)\zeta(6)+\frac{1}{(c\beta_e)^5}\int_0^{\infty} \frac{dx x^5}{(e^x-1)^2}\right)\nn
 &=&  2M_0(AT^5+BT_e^5),
 \eea 
where $ A= \frac{1}{c^5} \Gamma(5)\zeta(5),\, B=-\frac{1}{c^5} \Gamma(5)\zeta(5)+\frac{1}{c^5} \Gamma(6)\zeta(6)+\frac{1}{c^5}\int_0^{\infty} \frac{dx x^5}{(e^x-1)^2}$. For $\beta_e=\beta$, this result is identical to the GW result for scattering rate due to the electron phonon interaction and leads to the famously known Bloch formula for the resistivity in an equilibrium electron-phonon system\cite{Ziman}.\\
At a finite but low frequency the frequency dependent terms in Eqn. \ref{eq:case1} contributes as follows,
\bea
\Delta M''(\w\rightarrow 0)&=& -2\w^2 M_0\int_0^{q_D} dq q^4 \left( n''(\beta_e,\w_q)+\w_q n'''(\beta_e,\w_q) \right) \nn
&=& -2\w^2 \beta_e^2 M_0\int_0^{q_D} dq q^4 \left( n(\beta_e,\w_q)+3 n^2(\beta_e,\w_q)+2n^3(\beta_e,\w_q)\right.\nn
&&\left. -\beta_e\w_q n(\beta_e,\w_q)- 7\beta_e\w_q n^2(\w_q)-12\beta_e\w_q n^3(\w_q)-6\beta_e\w_q n^4(\w_q) \right).
\eea 
The coefficient involves integrals with various powers of the Bose-Einstein distribution function. They can be estimated as follows,
\be 
I_k=\int_0^{q_D} dq q^4 \beta_e\w_q n^k(\w_q)=\frac{1}{(\beta_e c)^5}\int_0^{\beta_e c q_D}  \frac{x^5}{(e^x-1)^k}dx\approx 
  \begin{cases}
  \frac{1}{(\beta_e c)^5}\int_0^{\infty}  \frac{x^5}{(e^x-1)^k}dx \sim T_e^5     & \quad \text{if } T_e<<\omega_D\\
    \frac{1}{(\beta_e c)^5}\int_0^{\beta_e c q_D}  x^{5-k} dx \sim T_e^{k-1}     & \quad \text{if } T_e>>\omega_D.\\
  \end{cases}
  \label{eq:Ik}
\ee
Thus, there is a $\w^2$ contribution and the coefficient has a $\beta_e^2 I_k$ dependence which $\sim T_e^3$ at the lowest order  for $T, T_e<<\w_D$, i.e. when both the temperatures are less than the Debye temperature. The $\w^2$ dependence shows that the Landau quasi-particle nature of the conducting electrons are preserved, as expected in case of weak electron-phonon interaction. \\
 Subcase (b): In the opposite limit, namely $T, T_e>\w_D$, i.e. when  both the electron temperature and the phonon temperature are higher than the Debye frequency, Eqn. \ref{eq:case1} takes the following form,
 \bea
 M''(\w=0)&=&2M_0\left( \frac{1}{(c\beta)^5}  \frac{(\beta c q_D)^4}{4}D_4(\beta c q_D)-\frac{1}{(c\beta_e)^5} \frac{(\beta_e c q_D)^4}{4}D_4(\beta_e c q_D)\right.\nn
 &&\left. +\frac{1}{(c\beta_e)^5} \frac{(\beta_e c q_D)^5}{5}D_5(\beta_e c q_D)+\frac{1}{(c\beta_e)^5}\int_0^{\beta_e c q_D} dx x^5\frac{1}{(e^x-1)^2}\right)\nn
 &\approx&2M_0\left( \frac{1}{c\beta}  \frac{ q_D^4}{4}-\frac{1}{ c\beta_e } \frac{  q_D ^4}{4}   +  \frac{ q_D}{5} +\frac{1}{(c\beta_e)^5} \frac{(\beta_e c q_D)^4}{4}\right)\nn
 &=&2M_0\left( \frac{q_D}{5} +\frac{1}{c\beta}  \frac{ q_D^4}{4}-\frac{1}{ c\beta_e } \frac{  q_D ^4}{4} +\frac{1}{ c\beta_e } \frac{  q_D^4}{4}\right)=2M_0(A_1+B_1T ).
 \eea
 Here $A_1=\frac{q_D}{5},\, B_1=\frac{1}{c}  \frac{ q_D^4}{4}$.  It is important to note that the scattering rate is independent of the electron temperature and is dictated by the phonon temperature only. \\ 
 If we consider the non-zero low frequency case, the frequency dependence in $\Delta M''(\w\rightarrow 0)$ is still $\sim \w^2$. But the coefficient has a linear in $T_e$ dependence.\\
 Case II: For frequencies greater than Debye frequencies, i.e. $\w>>\w_D$, the scattering rate given by the Eqn. \ref{eq:case1} takes the following form
 \bea 
 M''(\w)&=&M_0\int_0^{q_D} dq q^4\left( \left(\frac{1}{e^{\beta\w_q}-1}-\frac{1}{e^{-\beta_e\w}-1}\right)
 + \w\rightarrow -\w\right) \nn
 &=& - M_0\left(\frac{q_D^5}{5}\frac{1}{e^{-\beta_e\w}-1}- \int_0^{q_D} dq q^4\frac{1}{e^{\beta\w_q}-1}\right)  + \w\rightarrow -\w\nn
 &=& - M_0\left(\frac{q_D^5}{5}\frac{1}{e^{-\beta_e\w}-1}-\frac{1}{(c\beta)^5} \int_0^{\beta c q_D} dx x^4\frac{1}{e^{x}-1}\right)  + \w\rightarrow -\w ,\,\,\, x=\beta\omega_q=\beta c q\nn
 &=& - M_0\left(\frac{q_D^5}{5}\frac{1}{e^{-\beta_e\w}-1}- \frac{1}{(c\beta)^5} \frac{(\beta c q_D)^4}{4}D_4(\beta c q_D)\right)  + \w\rightarrow -\w \nn
 &=& - M_0\left(\frac{q_D^5}{5}\frac{1}{e^{-\beta_e\w}-1}- \frac{q_D^4}{4c\beta} D_4(\beta c q_D)\right)  + \w\rightarrow -\w .
 \eea
 \ewt
 If phonon temperature is much lower than the Debye frequency, i.e. for $\beta c q_D>>1$, the above equation reduces to,
 \bea 
  M''(\w)&=& -M_0\left(\frac{q_D^5}{5}\frac{1}{e^{-\beta_e\w}-1}- \frac{1}{(c\beta)^5} \Gamma(5)\zeta(5)\right) \nn
  && + (\w\rightarrow -\w) \nn
  &=& -M_0\left(\frac{q_D^5}{5}\frac{ 1}{e^{-\beta_e\w}-1}- \frac{1}{(c\beta)^5} \Gamma(5)\zeta(5)\right)\nn
  && -M_0\left(\frac{q_D^5}{5}\frac{1}{e^{\beta_e\w}-1}- \frac{1}{(c\beta)^5} \Gamma(5)\zeta(5)\right)\nn
  &=& M_0\left(\frac{q_D^5}{5}+\frac{2}{(c\beta)^5} \Gamma(5)\zeta(5)\right).
 \eea 
 For phonon frequencies greater than the Debye energy, i.e. for $\beta c q_D<<1$, the constant part of the above expression will remain same and the later part will $\sim T$ instead of $T^5$. It also notable that in  both cases $ M''(\w)$ is independent of the electron temperature. \\
Subcase c):  $T<< \w_D,\, T_e\rightarrow \infty$. 
The frequency dependence will be similar to the previous case. But the frequency dependent part will have $T^5$ dependence.
Thus we get various limiting expressions which are summarized in a table below. 
\bwt
\begin{center}
\begin{tabular}{|l|l|}
\hline
	$\w$ and $T$ regime & $M''(\w)$\\
\hline
\hline
	$\w= 0,\,\, T_e, T<<T_D$ & $2M_0(AT^5+BT_e^5 ),\,\, A=\frac{1}{c^5} \Gamma(5)\zeta(5),\, B=\frac{1}{c^5}\left(- \Gamma(5)\zeta(5)+ \Gamma(6)\zeta(6)+\int_0^{\infty} dx x^5\frac{1}{(e^x-1)^2}\right).$\\
\hline
	$\w= 0,\,\, T_e, T> T_D$ &$ 2M_0(A_1+B_1T),\, A_1=\frac{q_D}{5},\, B_1=\frac{1}{c}  \frac{ q_D^4}{4}$. \\
\hline
	$\w>>\w_D,\,\, T<< T_D$ & $ M_0\left(\frac{q_D^5}{5}+\frac{2T^5}{c} \Gamma(5)\zeta(5)\right)$\\
\hline
	$\w>>\w_D,\,\, T> T_D$ & $ M_0\left(\frac{q_D^5}{5}-T \frac{ q_D^4}{4c}\right)$\\
\hline
	$\w<< \w_D,\, T,\, T_e<T_D$ & $ 2M_0\left(T^5\frac{\Gamma(5)\zeta(5)}{c^5}+T_e^5\frac{\Gamma(5)}{c^5} + {\rm Const. }\times T_e^5\w^2\right)$\\
\hline
	$\w<< \w_D,\, T,\, T_e>T_D$ & $ 2M_0\left(\frac{q_D^4}{4c} T+{\rm Const. }\times T_e \w^2\right)$\\
\hline
	$\w<< \w_D,\, T<\w_D,\, T_e>T_D$ & $ 2M_0\left(T^5\frac{\Gamma(5)\zeta(5)}{c^5}+{\rm Const. }\times T_e \w^2\right)$\\
\hline
\end{tabular}
\end{center}
\ewt
However to understand the full dc and frequency  dependent behavior,
we need to perform the integral in Eqn.\ref{eq:ImMf} numerically. We choose the Debye energy $\w_D=0.03eV = 348.14 K$ to perform the integral. Now we present our numerical results. Possible reasons of various findings will be discussed in the next sections.
\figImMT\\
In Fig.\ref{fig:figimmt} we show the temperature dependence of the $M''(\w\rightarrow 0, T, T_e)$, normalized by
$M''(\w\rightarrow 0, T_D, T_D)$, which shows a monotonic increase with temperature. However there is a change in the curvature at low temperature as the electron temperature differs from that of the phonons. 
\figImM
\figImMmT
\figImMHT
 In Fig.\ref{fig:figimm},  plots of  $M''(\w, T, T_e)$ normalized by $M''(0, T, T_e)$ at a low electron temperature is shown. In this case we choose phonon temperature to be $0.001eV\sim 12K$. Here we see an increase followed by a trend of saturation with the increasing frequency. Also with higher electron temperature, electrons tend to decouple from the phonon bath and the magnitude of the scattering rate reduces. Fig.\ref{fig:figimm-mT} shows the same plot at an intermediate temperature 120K. Here we see much less rapid variation of scattering rate
with frequency compared to that observed in Fig.\ref{fig:figimm} near the Debye frequency.  Difference in the scattering rates  also reduces as the phonon temperature becomes higher. Except the change in the magnitude, same trend continues   at a temperature much higher than the Debye temperature as shown in Fig.\ref{fig:figimm-HT}.
\section{Summary and Conclusions}
\label{sec:discussion}
In Summary, we propose a generalization of the studies in electronic transport by memory function formalism to the non-equilibrium
situation. Here an electron-phonon coupled system is studied where the temperature of the electrons and the phonons are different,
as happens in many experimental situations. As explained in the previous sections, vertex corrections and the frequency dependence of the transport life time are also included in this approach naturally.
Thus this approach includes two subtle effects, namely some non-equilibrium effects and the proper account of the transport life time or the transport scattering rates of the electronic quasi-particles. This is quite timely and much sought after in context of recently flourishing non-equilibrium spectroscopic experiments.
In our results we see  1) in Fig.\ref{fig:figimmt}, if both the electron and the phonon temperatures are same, the dc scattering rate has a $T^5$ rise below the Debye temperature which changes to the linear in $T$ behavior in the high temperature.  This is in accord with the Bloch-Boltzmann formula and is a benchmark for our formalism in the equilibrium limit. 2) As the electron and the phonon temperatures start differing, in the same figure, we see a change in the curvature 
of the scattering rate below the Debye temperature. At higher electron temperature, due to the reduced Pauli blocking, more electronic states becomes available below the Fermi energy and the number of scattering events enhances. As a result scattering rate increases. 3) As we go above the Debye temperature, both the phonon generation and thus the electron-phonon scattering gets saturated. Thus the effects of the higher electron temperature or the non-equilibrium  effects get reduced. In this limit the difference in scattering rates at different electron temperatures becomes small.
4) In frequency dependent cases, we see difference in the electron and the phonon temperature leads to no qualitative change in the curve. But there is an order of magnitude change in the normalized values of the scattering rate at low temperature. The difference in magnitudes of the scattering rates at two different electron temperatures reduces as we go to the higher temperature which is also due to the saturation of phonon generation as explained previously. These are very much interesting findings from the present work. There are quite a few scopes of extensions on this work by considering, coupling with optical phonons, proper band structure for quasi-particles, electron-electron, electron-impurity interactions, external magnetic field (to study magneto-resistance and the non-equilibrium Hall effects), etc. and also in  more interesting non-equilibrium steady state situations. These are left for  future studies. 
This work can be treated as a basis for studying the effects of general electron-Boson coupling\cite{Basov} as well as non-equilibrium dynamics\cite{Zhang} often discussed in the case of cuprates and other strongly correlated systems.
\section*{Acknowledgement}
We thank Pankaj Bhalla for many useful discussions. 
\appendix
\bwt
\section{Derivation of Eqn. \ref{eq:ImMf}}
\label{sec:caln}
Since we are considering metals having degenerate electrons, the Fermi energy and the magnitude of Fermi wave vector $k_F$ are very large. For low energy scattering, angular changes in the scattering wave vectors are dominating while magnitude of the scattering wave vectors can be assumed constant and $\approx k_F$. Thus the $\theta$-integral can be considered separately and can be performed as follows,
\bea 
I_\theta&=&\int_0^\pi d\theta\sin\theta\delta(q-\sqrt{k^2+k'^2-2kk'\cos\theta})\nn
&\approx& \int_0^\pi d\theta\sin\theta\delta(q-k_F\sqrt{2(1-\cos\theta)})\nn
&=& \int_0^2 dx \delta(q-k_F\sqrt{2x})=\int_0^{2k_F} \frac{y}{k_F^2} dy\delta(q-y)=\frac{q}{k_F^2}.
\eea 
Considering $\chi_0=N_e/m$, and inserting the above expression for $I_\theta$, we get,
\bea
M''(\w)&=&\frac{2}{3}\pi\frac{ N}{N_e}\frac{1}{(2\pi)^4mm_ik_F^2}\int_0^\infty dq\frac{C_q^2}{\w_q}q^5\int_0^\infty k^2 dk\int_0^\infty k'^2 dk'\nn
&& \times\left\{ \frac{1}{\w}e^{\beta_e\epsilon_{k'}}\left( \frac{1}{e^{\beta_e\epsilon_k}+1}\frac{1}{e^{\beta_e\epsilon_{k'}}+1}\frac{1}{e^{\beta(\epsilon_k-\epsilon_{k'}+\w)}-1}\left(e^{\beta(\epsilon_k-\epsilon_{k'}+\w)}-e^{\beta_e(\epsilon_{k}-\epsilon_{k'})}\right) \right)\delta(\epsilon_k-\epsilon_{k'}-\omega_q+\w)\right.\nn
&& \left. + \w\rightarrow -\w\right \}.
\eea 
Next $k,\,k'$ integrals are there to be performed. We convert them as energy integrals as follows,
\bea 
\epsilon =\frac{k^2}{2m},\,\, d\epsilon=\frac{k}{m}dk=\sqrt{\frac{2\epsilon}{m}}dk\Rightarrow\int_0^\infty k^2 dk\int_0^\infty k'^2 dk'=2m^3\int_0^\infty\sqrt{\epsilon}d\epsilon\int_0^\infty\sqrt{\epsilon'}d\epsilon'\approx 2m^3 \epsilon_F\int_0^\infty d\epsilon\int_0^\infty d\epsilon'.
\eea 
Thus the imaginary part of the memory function can be rewritten in terms of the energy variables as,
\bea 
M''(\w)&=&\frac{4}{3}\pi\frac{ N}{N_e}\frac{m^2}{m_i}\frac{\epsilon_F}{(2\pi)^4k_F^2}\int_0^\infty dq\frac{C_q^2}{\w_q}q^5\int_0^\infty d\epsilon\int_0^\infty d\epsilon'\nn
&& \times\left\{ \frac{1}{\w}e^{\beta_e\epsilon'}\left( \frac{1}{e^{\beta_e\epsilon}+1}\frac{1}{e^{\beta_e\epsilon'}+1}\frac{1}{e^{\beta(\epsilon -\epsilon'+\w)}-1}\left(e^{\beta(\epsilon -\epsilon'+\w)}-e^{\beta_e(\epsilon-\epsilon' )}\right) \right)\delta(\epsilon -\epsilon'-\omega_q+\w)\right.\nn
&& \left. + \w\rightarrow -\w\right \}.
\eea 
After performing $\epsilon'$ integral  first, we get rid of the delta function and the resulting expression becomes,
\bea 
M''(\w)&=&\frac{4}{3}\pi\frac{ N}{N_e}\frac{m^2}{m_i}\frac{\epsilon_F}{(2\pi)^4k_F^2}\int_0^\infty dq\frac{C_q^2}{\w_q}q^5\int_0^\infty d\epsilon\nn
&& \times\left\{ \frac{1}{\w}e^{\beta_e(\epsilon -\omega_q+\w)}\left( \frac{1}{e^{\beta_e\epsilon}+1}\frac{1}{e^{\beta_e(\epsilon -\omega_q+\w)}+1}\frac{1}{e^{\beta\w_q}-1}\left(e^{\beta\w_q}-e^{\beta_e(\w_q-\w )}\right) \right)\right.\nn
&& \left. + \w\rightarrow -\w\right \}.
\eea
Now we are left with $\epsilon$ integral $I_\epsilon$, and is given as,
\bea 
I_\epsilon=\int_0^\infty d\epsilon \left(\frac{1}{e^{\beta_e\epsilon}+1} \right) \left(\frac{1}{e^{-\beta_e(\epsilon -\omega_q+\w)}+1} \right).
\eea 
The lower limit of the above integral can be considered as $-\infty$, when energy is measured with respect to the Fermi energy. Now we make the following substitution, 
\bea 
x=e^{\beta_e\epsilon},\,\, dx=\beta_e x d\epsilon.
\eea 
Using the above substitution $I_\epsilon$ can be rewritten and evaluated as,
\bea 
I_\epsilon&=&\int_0^\infty dx\frac{1}{\beta_ex} \left(\frac{1}{x+1} \right) \left(\frac{x e^{\beta_e( -\omega_q+\w)}}{x e^{\beta_e( -\omega_q+\w)}+1} \right)\nn
&=&\int_0^\infty dx\frac{\xi}{\beta_e} \left(\frac{1}{x+1}\right)\left(\frac{1}{x \xi +1} \right),\,\, \xi= e^{\beta_e( \w-\w_q)}\nn
&=&-\frac{1}{\beta_e} \frac{\xi}{\xi-1}\ln\left| \frac{x+1}{x+\frac{1}{\xi}}\right|_0^\infty
=\frac{1}{\beta_e} \frac{\xi}{\xi-1}\ln\left| \xi\right|= \frac{e^{\beta_e( \w-\w_q)}}{e^{\beta_e( \w-\w_q)}-1}( \w-\w_q).
\eea 
Thus the imaginary part of the memory function is left with only one integral over $q$ variable and takes the form,
\bea 
M''(\w)&=&\frac{4}{3}\pi\frac{ N}{N_e}\frac{m^2}{m_i}\frac{\epsilon_F}{(2\pi)^4k_F^2}\int_0^\infty dq\frac{C_q^2}{\w_q}q^5\nn
&& \times\left\{ \frac{1}{\w}\left(\frac{1}{e^{\beta\w_q}-1}\left(e^{\beta\w_q}-e^{\beta_e(\w_q-\w )}\right) \frac{e^{\beta_e( \w-\w_q)}}{e^{\beta_e( \w-\w_q)}-1}( \w-\w_q)\right)
 + \w\rightarrow -\w\right \}.
\eea 
Using the relations  $C_q=1/\rho_F,\,\, \w_q=cq$ and defining $M_0=\frac{4}{3}\pi\frac{ N}{N_e}\frac{m^2}{m_i}\frac{\epsilon_F}{(2\pi)^4k_F^2}\frac{1}{c\rho_F^2}$,
\bea 
M''(\w)&=& M_0\int_0^{q_D} dq q^4\left\{ \frac{1}{\w}\left(\frac{1}{e^{\beta\w_q}-1}\left(e^{\beta\w_q}-e^{\beta_e(\w_q-\w )}\right) \frac{e^{\beta_e( \w-\w_q)}}{e^{\beta_e( \w-\w_q)}-1}( \w-\w_q)\right)
 + \w\rightarrow -\w\right \}\nn
&=&  M_0\int_0^{q_D} dq q^4\left\{ \frac{1}{\w}\left(\frac{1}{e^{\beta\w_q}-1}\left(e^{\beta_e(\w_q-\w )}-e^{\beta\w_q}\right) \frac{1}{e^{\beta_e( \w_q-\w)}-1}( \w-\w_q)\right)
 + \w\rightarrow -\w\right \}\nn
 &=& M_0\int_0^{q_D} dq q^4\left\{ \frac{1}{\w}\left(\left(\frac{1}{e^{\beta\w_q}-1}-\frac{1}{e^{\beta_e( \w_q-\w)}-1}\right)  ( \w-\w_q)\right)
 + \w\rightarrow -\w\right \}\nn
 &=& M_0\int_0^{q_D} dq q^4\left\{ \frac{1}{\w}\left(\left(n(\beta,\w_q)-n(\beta_e, \w_q-\w)\right)  ( \w-\w_q)\right)
 + \w\rightarrow -\w\right \}.
 \label{eq:ImMf-app}
 \eea
\section{Derivatives of the Bose function}
\label{sec:dn}
Various derivatives of the Bose function can be written in terms of the Bose function itself as,
\bea 
n'(\w_q-\w)&=&\frac{\partial n(\w_q-\w)}{\partial (\w_q-\w)}=\frac{\partial }{\partial (\w_q-\w)}\left(\frac{1}{e^{\beta(\w_q-\w)}-1}\right)\nn
&=&-\beta\left(\frac{e^{\beta(\w_q-\w)}}{\left(e^{\beta(\w_q-\w)}-1\right)^2}\right)=-\beta\left(n(\w_q-\w)+n^2(\w_q-\w)\right),\\
n''(\w_q-\w)&=&-\beta\left(n'(\w_q-\w)+2n(\w_q-\w)n'(\w_q-\w)\right)\nn
&=&-\beta\left(1+2n(\w_q-\w)\right)\left(n(\w_q-\w)+n^2(\w_q-\w)\right)\nn
&=& \beta^2\left(n(\w_q-\w)+3n^2(\w_q-\w)+2n^3(\w_q-\w)\right),\\
n'''(\w_q-\w)&=&\beta^2\left(1+6n(\w_q-\w)+6n^2(\w_q-\w)\right)n'(\w_q-\w)\nn
&=&-\beta^3\left(1+6n(\w_q-\w)+6n^2(\w_q-\w)\right)\left(n(\w_q-\w)+n^2(\w_q-\w)\right)\nn
&=&-\beta^3\left(n(\w_q-\w)+7n^2(\w_q-\w)+12n^3(\w_q-\w)+6n^4(\w_q-\w)\right).
\eea 
\section{Debye functions}
\label{sec:d-func}
The first Debye function is defined as,
\bea 
D^1_n(x)=\int_0^x dt \frac{t^n}{e^t-1}=x^n\left(\frac{1}{n}-\frac{x}{2(n+1)}+\sum_{k=1}^\infty\frac{B_{2k}x^{2k}}{(2k+n)(2k)!}\right),
\eea 
for $|x|<<2\pi,\, n\ge 1$, and $B_{2k}$'s are Bernoulli numbers. The second Debye function is defined by,
\bea 
D^2_n(x)=\int^\infty_x dt \frac{t^n}{e^t-1}=\sum_{k=1}^\infty e^{-kx}\left(\frac{x^n}{n}+\frac{nx^{n-1}}{k^2}+\frac{n(n-1)x^{n-2}}{k^3}+\cdots+\frac{n!}{k^{n+1}}\right),
\eea 
for $x>0, n\ge 1$
Sum of the above two integrals,
\bea 
D^1_n(x)+D^2_n(x)=\int^\infty_0 dt \frac{t^n}{e^t-1}=n!\zeta(n+1)
\eea
The Riemann zeta function used in the above equation is given as,
\bea 
\zeta(x)=\frac{1}{\Gamma(x)}\int_0^\infty\frac{u^{x-1}}{e^u-1}du
\eea
where $\Gamma(x)$ is the gamma function. If $x$ is an integer $n$, 
\bea 
\zeta(n)&=&\frac{1}{\Gamma(n)}\int_0^\infty\frac{u^{n-1}}{e^u-1}du=\frac{1}{\Gamma(n)}\int_0^\infty\sum_{k=1}^\infty e^{-ku}u^{n-1}du=\sum_{k=1}^\infty\frac{1}{k^n}.
\eea
\ewt

\end{document}